\documentclass[runningheads]{llncs}
\usepackage[T1]{fontenc}
\usepackage{amsmath}
\usepackage{graphicx}
\usepackage{booktabs}
\begin{document}
\title{Application of Machine Learning Models for Carbon Monoxide and Nitrogen Oxides Emission Prediction in Gas Turbines}
\titlerunning{Machine Learning Models for CO and NOx Predictions}
%
\author{Kamyar Zeinalipour\inst{1} \and
Laure Barriere\inst{2} \and
David Ghelardi\inst{2} \and
Marco Gori\inst{1}
\authorrunning{Zeinalipour et al.}
%
\institute{University of Siena, Department Of Information Engineering And Mathematics, via Roma, 56, 53100 Siena, Italy \\
\email{\{kamyar.zeinalipour2, marco.gori\}@unisi.it}\\
 \and
Baker Hughes S.r.l., Piazza Enrico Mattei, 50127 Firenze, Italy\\
\email{\{laure.barriere, david.ghelardi\}@bakerhughes.com}}}
\maketitle              
\begin{abstract}
This paper addresses the 
environmental impacts linked to hazardous emissions from gas turbines, with a specific focus on employing various machine learning (ML) models to predict the emissions of Carbon Monoxide (CO) and Nitrogen Oxides (NOx) as part of a Predictive Emission Monitoring System (PEMS). We employ a comprehensive approach using multiple predictive models to offer insights on enhancing regulatory compliance and optimizing operational parameters to reduce environmental effects effectively. Our investigation explores a range of machine learning models including linear models, ensemble methods, and neural networks. The models we assess include Linear Regression, Support Vector Machines (SVM), Decision Trees, XGBoost, Multi-Layer Perceptron (MLP), Long Short-Term Memory networks (LSTM), Gated Recurrent Units (GRU), and K-Nearest Neighbors (KNN). This analysis provides a comparative overview of the performance of these ML models in estimating CO and NOx emissions from gas turbines, aiming to highlight the most effective techniques for this critical task. Accurate ML models for predicting gas turbine emissions help reduce environmental impact by enabling real-time adjustments and supporting effective emission control strategies, thus promoting sustainability.
\end{abstract}

\section{Introduction}
\begin{figure}[ht]
    \centering
    \includegraphics[width=0.7\textwidth]{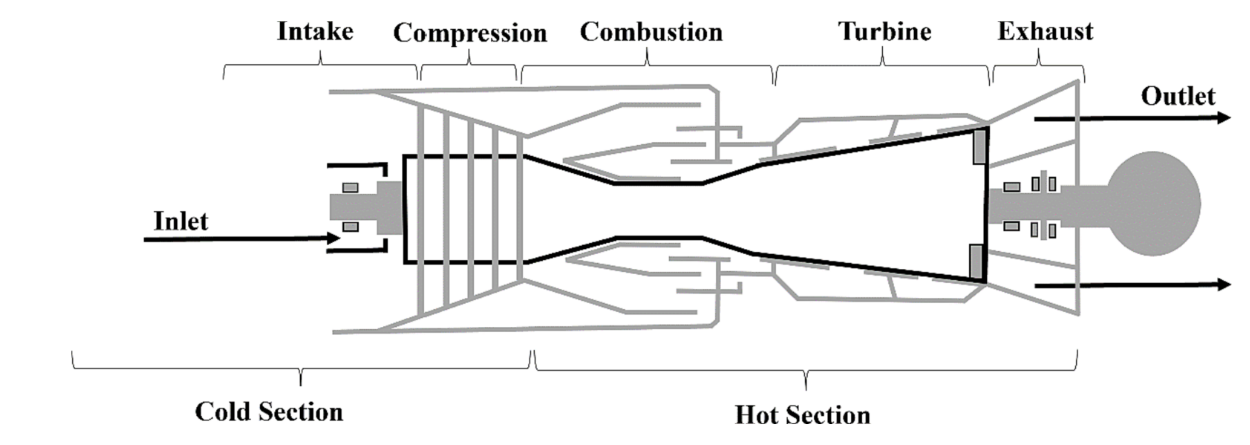}
    \caption{A typical industrial gas turbine used in power generation.}
    \label{fig:gasturbine}
\end{figure}

The 
increasing concerns about environmental degradation necessitate a sustained focus on reducing pollutants and emissions that significantly contribute to global warming and poor air quality. Gas turbines, widely used in power generation and aviation industries, are notable sources of hazardous emissions, including Carbon Monoxide (CO) and Nitrogen oxide (NOx). These emissions pose severe risks to both environmental health and human well-being, highlighting an urgent need for better regulatory compliance and operational optimization. In the past, most traditional methods for controlling and predicting emissions depended on static regulatory mechanisms and after-the-fact mitigation strategies. However, with the advent of advanced analytics and machine learning (ML) technologies, dynamic and proactive solutions are now feasible.\\
This paper explores the application of various machine learning models to predict and analyze the emissions from gas turbines, critically focusing on CO and NOx. Anticipating emission levels accurately not only helps in adhering to environmental regulations more effectively but also assists in enhancing the operational efficiency of gas turbines. Our research delves into a comprehensive suite of machine learning techniques ranging from basic linear models to more sophisticated neural networks. Specifically, we investigate Linear Regression, Support Vector Machines (SVM), Decision Trees, XGBoost, Multi-Layer Perceptron (MLP), Long Short-Term Memory networks (LSTM), Gated Recurrent Units (GRU), and K-Nearest Neighbors (KNN). Each of these models offers unique strengths and limitations in handling the complexity and variability of emission data from gas turbines. Figure \ref{fig:gasturbine} showcases a typical gas turbine, highlighting the intricate machinery involved and underscoring the importance of precise emission monitoring and control.

The comparative analysis conducted in this study aims to identify which ML models not only predict emissions most accurately but also adapt effectively to real-time data, thereby providing actionable insights that can influence immediate operational decisions. This could improve how industries approach emission reduction, shifting from reactive to preemptive strategies. Through a rigorous evaluation of these models, the paper seeks to contribute to the body of knowledge in environmental management by outlining effective ML-driven approaches for enhancing emission controls, ultimately aiding the global pursuit of sustainability and environmental protection.
The composition of this document is methodically organized into several key sections: a literature review in Section~\ref{sec:relatedworks}, an outline of the methodology in Section~\ref{sec:Methodology}, an analysis of the experiments in Section~\ref{sec:Experiments}, and concluding remarks in Section~\ref{sec:conclusions}.
\section{Related Work}\label{sec:relatedworks}

This paper synthesizes diverse machine learning approaches for Predictive Emission Monitoring Systems (PEMS), showcasing an evolution through multimodal frameworks. Introduced by \cite{si2019development}, these methodologies span several foundational approaches:
\begin{enumerate}
    \item \textbf{Mechanistic Models:} Harnessing principles of mechanics spanning thermodynamics and kinetics to derive relationships through physical equations, establishing a traditional but robust foundation \cite{hung1975experimentally}.
    \item \textbf{Statistical Models:} Harnessing correlations and tendencies within operational data to predict emissions, occasionally incorporating the dynamical inputs from mechanistic frameworks to form what is often termed hybrid models \cite{lee2005application,saiepour2006development}.
    \item \textbf{Computational Intelligence:} Leveraging modern computational paradigms, including but not limited to neural networks, decision trees, and genetic algorithms, this approach is characterized by its adaptability and potential for handling complex nonlinear relationships \cite{kaya2019predicting,vanderhaegen2010predictive,chawathe2021explainable,Kochueva2021explainable}.
\end{enumerate}
Key milestones in the domain of PEMS commence with the year 1973, marking the segment's initiation through the efforts of integrating thermodynamics with predictive monitoring \cite{hung1975experimentally}. Further innovation is observed in the work from 1993 where configurations featuring multiple synthetic reactors were used to model varied combustive zones \cite{rizk1993semianalytical}.\\
By the year 2003, advancements took a theoretical turn with regression and combustion theory collaborating to facilitate emission predictions \cite{chien2003feasibility}, transitioning in 2005 towards employing multivariate statistical methods to address emission challenges in heater systems \cite{lee2005application}.\\
The pivotal year of 2010 directed the spotlight towards neural network models designed for resilience, primarily serving as fail-safes for conventional emission monitoring systems \cite{vanderhaegen2010predictive}. Accelerated interest in machine learning from 2019 led to a transparent dialogue around dataset characteristics and model efficacy, fueled by the accessibility of datasets such as those from University of California, Irvine \cite{kaya2019predicting}.\\
Progressing into the years 2020 to 2023, exploration of diverse algorithmic strategies such as K-nearest-neighbor, gradient boosting, and novel ensemble models like Deep Forest indicated a new era of computational intelligence in PEMS \cite{rezazadeh2020environmental,si2020development,Coelho2024DeepForest}.
Our investigation aligns with this trajectory by evaluating an array of contemporary models against specific challenges pertinent to CO and NOx emissions in predefined gas turbine configurations, while also addressing issues emerging from shifts in data distribution.
\section{Methodology}\label{sec:Methodology}
The methodology of this research involves a structured approach to predict the levels of Carbon Monoxide (CO) and Nitrogen Oxides (NOx) emissions from gas turbines using various machine learning models. The objective is to evaluate each model’s efficacy in terms of accuracy and reliability for inclusion in a Predictive Emission Monitoring System (PEMS). Each model was trained and tested using historical emission data obtained from gas turbine systems, ensuring a robust dataset to train the predictive models. The models include both traditional algorithms and advanced machine learning techniques, described briefly in the following paragraphs.

\paragraph{Linear Regression (LR)} is a fundamental machine learning model applied to predict a continuous outcome variable (emissions) based on the linear relationship with predictor variables. For this study, LR was implemented to establish a baseline understanding of the emissions prediction. Variable selection was conducted to identify the most significant predictors, ensuring the model focuses on factors directly impacting CO and NOx emissions.
\paragraph{Support Vector Machines (SVM)} was utilized for its capacity to handle non-linear relationships through the employment of different kernels. In this analysis, both linear and radial basis function (RBF) kernels were tested to study emission predictions under linear and non-linear assumptions, respectively. SVMs are particularly noted for their effectiveness in high-dimensional spaces and their robustness against overfitting, especially in complex environments like emission data.
\paragraph{Decision Trees} model was employed to facilitate interpretability of the results. It uses a tree-like graph of decisions and their possible consequences, including chance event outcomes. This model is beneficial for understanding the decision rules that the system might be considering, and for visualizing the paths and splits in key variables that lead to higher emissions.
\paragraph{XGBoost} was applied for its superior performance in handling outliers, missing values, and numerous features in the dataset. XGBoost is known for its speed and performance, primarily due to its tree pruning and handling of sparse data. This technique combines multiple weak predictive models to build a robust ensemble model conducive to capturing complex patterns in data.
\paragraph{K-Nearest Neighbors (KNN)}
was used to investigate its suitability for this application due to its simplicity and effectiveness in classification problems. It was hypothesized that emission levels could be predicted based on the similarity of operational parameters in past cases, thereby creating clusters of similar operational scenarios to predict the emission outputs.
\paragraph{Multi-Layer Perceptron (MLP)}
 a type of neural network, was tested for its ability to learn non-linear models. It consists of multiple layers wherein each layer is fully connected to the next layer. The MLP was trained with various architectures and activation functions to find the best structure for predicting emissions effectively.
\paragraph{Long Short-Term Memory networks (LSTM) and Gated Recurrent Units (GRU)}
Both LSTM and GRU models are types of recurrent neural networks that are particularly useful for modeling sequence data. Given the time-dependent characteristics of emission data, these models were evaluated to see how well they can predict future emissions based on past patterns. They are particularly adept at learning dependencies and relationships in time-series data.

\subsection{Evaluation}
To ensure the precision and reliability of the predictive models developed for estimating CO and NOx emissions from gas turbines, a robust evaluation framework utilizing several statistical metrics is crucial. This approach is fundamental for verifying the accuracy of the model predictions against actual observed values.\\
The selected metrics below, commonly used in predictive modeling, help quantify the model's performance by measuring discrepancies between predicted and observed values:

\paragraph{Root Mean Square Error (RMSE)}
\begin{equation}
\text{RMSE} = \sqrt{\frac{1}{n} \sum_{i=1}^{n} (\hat{y}_i - y_i)^2}
\end{equation}
RMSE calculates the model's average prediction errors, providing insight into their typical magnitude.

\paragraph{Mean Absolute Percentage Error (MAPE)}
\begin{equation}
\text{MAPE} = \left( \frac{1}{n} \sum_{i=1}^{n} \left|\frac{\hat{y}_i - y_i}{y_i}\right| \right) \times 100\%
\end{equation}
MAPE expresses accuracy as a percentage, useful for assessing errors relative to actual values' scale.

\paragraph{Mean Squared Error (MSE)}
\begin{equation}
\text{MSE} = \frac{1}{n}\sum_{i=1}^{n} (\hat{y}_i - y_i)^2
\end{equation}
MSE emphasizes and penalizes larger errors, beneficial when significant errors are particularly undesirable.

\paragraph{Mean Absolute Error (MAE)}
\begin{equation}
\text{MAE} = \frac{1}{n}\sum_{i=1}^{n} |\hat{y}_i - y_i|
\end{equation}
MAE provides a straightforward average of error magnitudes, easy to interpret and valuable for assessing overall model accuracy.

These metrics collectively offer a comprehensive view of model performance, addressing both the magnitude and accuracy of errors, crucial for effective emissions control strategies in gas turbines.
\section{Experiments}\label{sec:Experiments}
In this section, we will explore the experiments conducted for this study, starting with an overview of the dataset utilized in the project. Following that, we will delve into the various model architectures and the hyperparameters employed. Finally, we will discuss the results obtained from these experiments.
\subsection{Data Collection}
Our research employed a dataset compiled from August 2019 to April 2021, with 113,675 data entries recorded under ABC mode, indicating steady-state full-load operations of the gas turbine for consistent measurements. The dataset contains 13 numerical variables, such as pressure, temperature, and flow rate, facilitating computational analysis.\\
In preparation for analysis using various models, distinct preprocessing steps were taken. We standardized non-XGBoost model features by removing the mean and scaling to unit variance via:
\[
z = \frac{(x - u)}{s}
\]
where \( x \) is the original feature value, \( u \) is the mean, and \( s \) is the standard deviation across samples. This normalization aids model convergence by stabilizing input magnitudes.\\
Specifically for LSTM, a windowing technique partitioned the data to capture short-term temporal correlations crucial for sequential trend analysis. Conversely, XGBoost and decision tree used raw data due to its inherent insensitivity to feature scale variation, simplifying its preparation phase.\\
This tailored preprocessing approach ensures optimal data configuration for each model, leveraging their respective strengths.
\subsection{Training Setup}
In this section, we describe the training setup for various models employed in our study. The optimal hyperparameters and architectures were identified based on the Minimum Mean Squared Error (MSE) observed in the validation set.
\paragraph{Linear Regression}
For Linear Regression, we implemented Elastic Net regularization. Specifically, for predicting CO and NOx emissions, the parameters were set to $\alpha = 0.1$ and $\text{l1\_ratio} = 0.1$.
\paragraph{Support Vector Machine}
We employed the Support Vector Machine (SVM) with the following hyperparameters to predict the emissions of CO and NOx. For NOx prediction, we used $C=10$, $gamma= scale$, and $kernel = rbf$. For CO prediction, we used $C=100$, $gamma= scale$, and $kernel = rbf$.
\paragraph{Decision Trees}
We utilized the G to forecast CO and NOx emissions, configuring distinct hyperparameters for each. For predicting NOx, we specified: $max\_depth = None$, $min\_samples\_leaf = 4$, $min\_samples\_split = 2$, and $max\_features = \text{sqrt}$. Conversely, for CO prediction, the settings were: $max\_depth = 30$, $min\_samples\_leaf = 1$, $min\_samples\_split = 5$, and $max\_features = \text{sqrt}$.
\paragraph{XGBoost}
We deployed the XGBoost algorithm for predicting CO and NOx emissions, tailoring specific hyperparameters for each pollutant. For NOx predictions, we set the parameters as follows: $max\_depth = 8$, $learning\_rate= 0.1$, and $n\_estimators = 300$. In contrast, for CO forecasts, the configuration was slightly different: $max\_depth = 8$, $learning\_rate= 0.01$, and $n\_estimators = 300$.
\paragraph{Multi-Layer Perceptron}
We implemented a Multi-Layer Perceptron (MLP) architecture to predict CO and NOx emissions, customizing the hyperparameters specifically for each pollutant. The model configuration was consistent across both pollutants, utilizing a five-layer structure. Each of the first four layers comprises neurons in the following sequence: 256, 128, 64, and 32, all employing the ReLU activation function. The architecture concludes with a dense output layer of a single neuron using a linear activation function, suitable for addressing regression problems.
The hyperparameters were set as follows: the number of epochs ($num\_epochs$) was 100, learning rate ($learning\_rate$) was set to 0.01, and the batch size ($batch\_size$) was fixed at 64.
\paragraph{Long Short-Term Memory}
We implemented a Long Short-Term Memory (LSTM) model to forecast CO and NOx emissions, tailoring the hyperparameters specifically for each pollutant. The model maintains a consistent configuration for both pollutants, featuring a six-layer architecture. The initial two layers are LSTM layers with 64 and 32 neurons, respectively, followed by three dense layers with neuron counts arranged in the sequence: 128, 64, and 32. All layers employ the ReLU activation function. The architecture is rounded off with a dense output layer containing a single neuron using a linear activation function, which is ideal for regression tasks.
The hyperparameters included a training duration of 100 epochs ($num\_epochs$), a learning rate ($learning\_rate$) of 0.001, and a batch size ($batch\_size$) set to 64.
\paragraph{Gated Recurrent Unit}
For predicting CO and NOx emissions, we utilized a Gated Recurrent Unit (GRU) with hyperparameters customized for each pollutant. The model consists of six layers: two initial GRU layers with 64 and 32 neurons, followed by three dense layers configured in decreasing neuron counts of 128, 64, and 32, all using the ReLU activation function. The architecture concludes with a single-neuron dense output layer using a linear activation function, suitable for regression tasks.
The model operates over 100 epochs (\(\$num\_epochs\)), with a learning rate (\(\$learning\_rate\)) of 0.001 and batch size (\(\$batch\_size\)) of 64, ensuring effective learning and adaptation to pollution data.
\paragraph{K Nearest Neighbor Regressor}
The K Nearest Neighbor Regressor (KNN) algorithm was utilized to predict CO and NOx emissions, with tailored hyperparameters specific to each pollutant. For predicting NOx, the hyperparameters were set to: $n\_neighbors = 4$, $weights = \text{distance}$, and $algorithm = \text{brute}$. Conversely, the configuration for CO predictions differed slightly, employing: $n\_neighbors = 4$, $weights = \text{distance}$, and $algorithm = \text{ball\_tree}$.

\subsection{Results and  Discussion}
\begin{figure}[ht]
    \centering
    \includegraphics[width=0.7\textwidth]{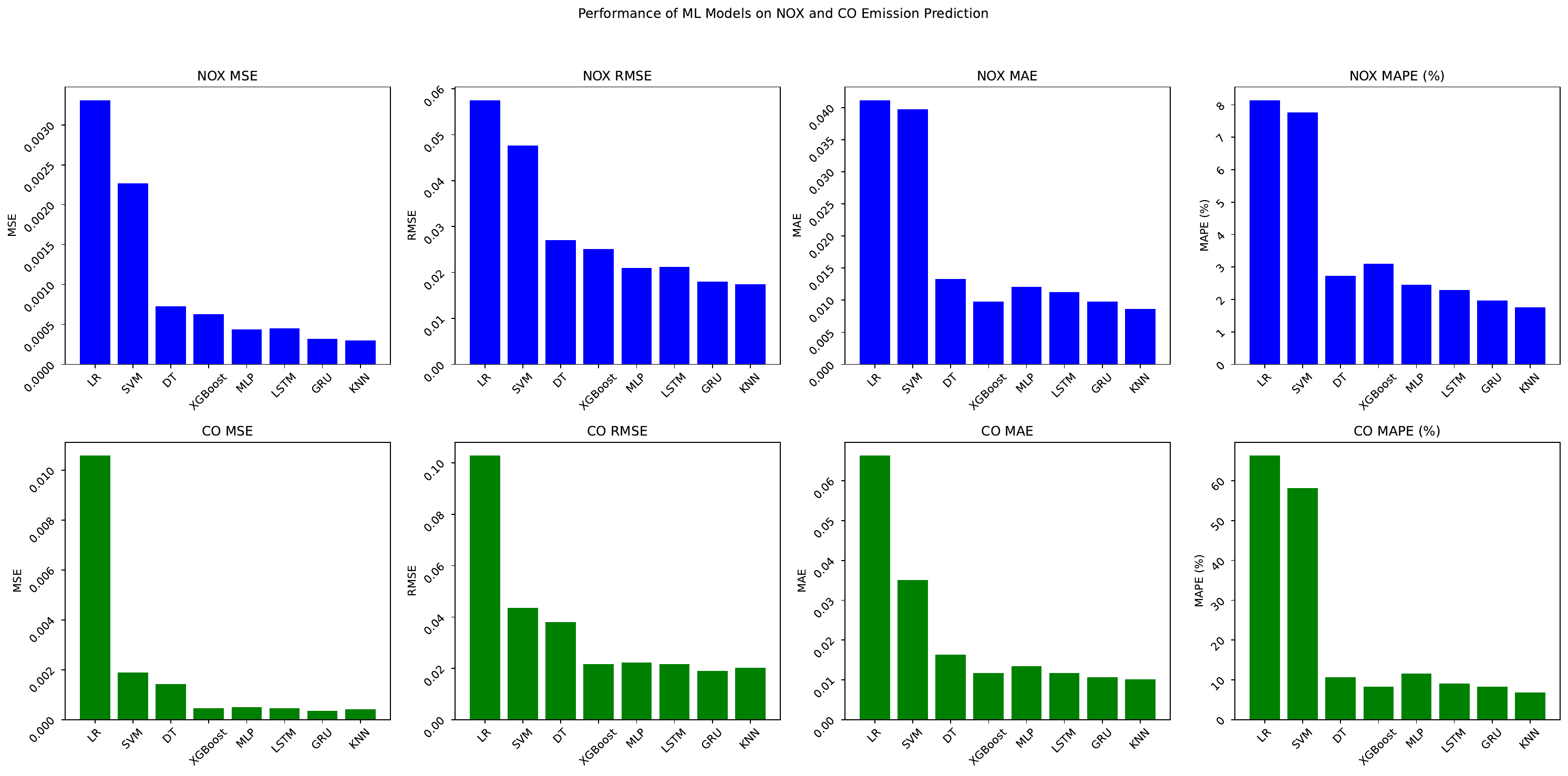}
    \caption{Normalized performance of various ML models on CO and NOx emission prediction.}
    \label{fig:performance}
\end{figure}

The analysis demonstrated different strengths among the machine learning models in forecasting emissions. Table \ref{tab:performance_nox_c} provide a summary of the normalized performance metrics for each model on the test set.
Additionally, Figure \ref{fig:performance} presents a histogram plot of all performance metrics—MSE, RMSE, MAE, and MAPE—for each model, facilitating a deeper understanding of the variance in model performances.

\begin{table}[h]
    \centering
    \caption{Normalized performance of various ML models on NOX and CO emission prediction.}
    \label{tab:performance_nox_co}
    \begin{tabular}{l|cccc|cccc}
        \toprule
        & \multicolumn{4}{c|}{Normalized NOX Emissions} & \multicolumn{4}{c}{Normalized CO Emissions} \\
        \cmidrule{2-9}
        ML Model & MSE & RMSE & MAE & MAPE (\%) & MSE & RMSE & MAE & MAPE (\%) \\
        \midrule
        Linear Regression & 0.00331 & 0.05752 & 0.04116 & 8.14289 & 0.01058 & 0.10284 & 0.06632 & 66.29603 \\
        SVM              & 0.00227 & 0.04769 & 0.03970 & 7.75883 & 0.00190 & 0.04359 & 0.03514 & 58.16424 \\
        Decision Trees   & 0.00073 & 0.02710 & 0.01333 & 2.72239 & 0.00144 & 0.03797 & 0.01635 & 10.61613 \\
        XGBoost          & 0.00063 & 0.02509 & 0.00975 & 3.09573 & 0.00047 & 0.02158 & 0.01168 & 8.30232 \\
        MLP              & 0.00044 & 0.02104 & 0.01204 & 2.44558 & 0.00050 & 0.02226 & 0.01343 & 11.62583 \\
        LSTM             & 0.00045 & 0.02118 & 0.01129 & 2.28486 & 0.00047 & 0.02177 & 0.01179 & 9.12560 \\
        GRU              & 0.00032 & 0.01798 & 0.00978 & 1.97100 & \textbf{0.00036 }& \textbf{0.01896 }& 0.01062 & 8.34426 \\
        KNN              & \textbf{0.00030} & \textbf{0.01741} & \textbf{0.00864} & \textbf{1.75548} & 0.00041 & 0.02033 & \textbf{0.01020} & \textbf{6.90689} \\
        \bottomrule
    \end{tabular}
    \label{tab:performance_nox_c}
\end{table}

The results from machine learning models on predicting NOx and CO emissions from gas turbines provide insightful comparisons regarding model efficiency and accuracy. When assessing the models on NOx emission predictions, there is a considerable variation in performance metrics including Root Mean Square Error (RMSE), Mean Absolute Error (MAE), and Mean Absolute Percentage Error (MAPE). The K-Nearest Neighbors (KNN) model outshines other models consistently across all metrics, achieving the lowest RMSE, MAE, and MAPE values of 0.00030, 0.00864, and 1.75548\% respectively, which indicates its superior predictive capability with minimal prediction errors and variability in estimating NOx emissions. Other models such as the Gated Recurrent Unit (GRU) and Long Short-Term Memory (LSTM) networks also demonstrate commendable results, especially in terms of MAPE, suggesting effective handling of time-series data from turbines, albeit slightly lagging behind KNN in overall performance.\\
On the other hand, the analysis of CO emissions predictions reveals a similar trend where the GRU achieves the lowest MSE and RMSE (0.00036 and 0.01896 respectively), yet KNN leads with the smallest MAE and MAPE (0.01020 and 6.90689\% respectively), reinforcing its efficacy across both pollutants. Models such as XGBoost, Multilayer Perceptron (MLP), and LSTM show better accuracy than more traditional approaches like Linear Regression and Support Vector Machines (SVM), which report significantly higher error rates across all metrics. Notably, Linear Regression demonstrates considerable inadequacy in its predictive capability for CO emissions, as indicated by a particularly high MAPE of 66.29603\%. This stark difference highlights the complex nature of emission predictions that benefit from more sophisticated, non-linear models capable of capturing intricate patterns in the data produced by gas turbines.

\section{Conclusion}\label{sec:conclusions}
In conclusion, this paper demonstrates the efficacy of machine learning (ML) models in predicting hazardous emissions from gas turbines, focusing on Carbon Monoxide (CO) and Nitrogen Oxides (NOx). Our analysis highlighted the superior performance of the K-Nearest Neighbors (KNN) model in accurately forecasting emissions, making it highly suitable for integration into Predictive Emission Monitoring Systems (PEMS). Gated Recurrent Units (GRU), Long Short-Term Memory networks (LSTM), and XGBoost also showed promising results, though KNN consistently outperformed other models.\\
Our findings underscore the limitations of traditional linear models like Linear Regression in handling complex emission data, advocating for the utilization of more sophisticated, non-linear models. The strategic application of effective ML models can enhance regulatory compliance and facilitate real-time operational adjustments, thereby reducing environmental impacts and advancing sustainability in industrial practices. This research encourages ongoing enhancements and widespread adoption of advanced ML techniques for improved emission control and monitoring.


\begin{thebibliography}{00}
\bibitem{si2019development}Si, M., Tarnoczi, T., Wiens, B. \& Du, K. Development of predictive emissions monitoring system using open source machine learning library–Keras: A case study on a cogeneration unit. {\em IEEE Access}. \textbf{7} pp. 113463-113475 (2019)
\bibitem{hung1975experimentally}Hung, W. An experimentally verified NOx emission model for gas turbine combustors. {\em Turbo Expo: Power For Land, Sea, And Air}. \textbf{79771} pp. V01BT02A009 (1975)
\bibitem{rizk1993semianalytical}Rizk, N. \& Mongia, H. Semianalytical correlations for NOx, CO, and UHC emissions.  (1993)
\bibitem{chien2003feasibility}Chien, T., Chu, H., Hsu, W., Tseng, T., Hsu, C. \& Chen, K. A feasibility study on the predictive emission monitoring system applied to the Hsinta power plant of Taiwan Power Company. {\em A\&WMA \textbf{53}, 1022-1028 (2003)}
\bibitem{lee2005application}Lee, Y., Kim, M. \& Han, C. Application of multivariate statistical models to prediction of NO x emissions from complex industrial heater systems. {\em Journal Of Environmental Engineering}. \textbf{131}, 961-970 (2005)
\bibitem{saiepour2006development}Saiepour, M., Schofield, N., Leden, B., Niska, J., Link, N., Unamuno, I. \& Gomes, J. Development and assessment of predictive emission monitoring systems (PEMS) models in the steel industry. {\em Proc. Iron Steel Technol. Conf.}. \textbf{2} pp. 1121-1132 (2006)
\bibitem{vanderhaegen2010predictive}Vanderhaegen, E., Deneve, M., Laget, H., Faniel, N. \& Mertens, J. Predictive emissions monitoring using a continuously updating neural network. {\em Turbo Expo: Power For Land, Sea, And Air}. \textbf{43970} pp. 769-775 (2010)
\bibitem{kaya2019predicting}Kaya, H., Tüfekci, P., Uzun, E. \& Others Predicting CO and NOxemissions from gas turbines: novel data and abenchmark PEMS.. {\em TURKISH J. Electr. Eng. Comput. Sci.}. \textbf{27}, 4783-4796 (2019)
\bibitem{rezazadeh2020environmental}Rezazadeh, A. Environmental Pollution Prediction of NOx by Process Analysis and Predictive Modelling in Natural Gas Turbine Power Plants. {\em ArXiv Preprint ArXiv:2011.08978}. (2020)
\bibitem{si2020development}Si, M. \& Du, K. Development of a predictive emissions model using a gradient boosting machine learning method. {\em Environmental Technology \& Innovation}. \textbf{20} pp. 101028 (2020)
\bibitem{chawathe2021explainable}Chawathe, S. Explainable Predictions of Industrial Emissions. {\em 2021 IEEE International IOT, Electronics And Mechatronics Conference (IEMTRONICS)}. pp. 1-7 (2021)
\bibitem{Kochueva2021explainable}Kochueva, O. \&  Nikolskii K. Data Analysis and Symbolic Regression Models for Predicting CO and NOx Emissions from Gas Turbines  {\em  Computation vol. 9, no. 12, p. 139, Dec. 2021} 
\bibitem{Coelho2024DeepForest}dos Santos Coelho, L., Ayala, H. V. H. \& Mariani, V. C. CO and NOx emissions prediction in gas turbine using a novel modeling pipeline based on the combination of deep forest regressor and feature engineering
 {\em  Fuel vol. 355, p.129366, Jan. 2024} 

\bibitem{zou2005regularization}Zou, H. \& Hastie, T. Regularization and variable selection via the elastic net. {\em JRSS-B}. \textbf{67}, 301-320 (2005)
\bibitem{chen2016xgboost}Chen, T. \& Guestrin, C. Xgboost: A scalable tree boosting system. {\em Proceedings Of The 22nd Acm Sigkdd International Conference On Knowledge Discovery And Data Mining}. pp. 785-794 (2016)
\bibitem{article}Hochreiter, S. \& Schmidhuber, J. Long Short-term Memory. {\em Neural Computation}. \textbf{9} pp. 1735-80 (1997,12)
\bibitem{rumelhart1986learning}Rumelhart, D., Hinton, G. \& Williams, R. Learning representations by back-propagating errors. {\em Nature}. \textbf{323}, 533-536 (1986)
\bibitem{jolliffe2016principal}Jolliffe, I. \& Cadima, J. Principal component analysis: a review and recent developments. {\em Phil. Trans. R. Soc. A}. \textbf{374}, 20150202 (2016)


\end{thebibliography}
\end{document}